\documentclass{PoS}

\usepackage{calc}

\newcommand{\mrm}{\mathrm}

\newcommand{\bea}{\begin{eqnarray}}
\newcommand{\eea}{\end{eqnarray}}
\newcommand{\beq}{\begin{equation}}
\newcommand{\eeq}{\end{equation}}
\newcommand{\ec}{\end{center}}
\newcommand{\bc}{\begin{center}}
\newcommand{\gev}{{\rm GeV}}

\newcommand{\pdir}{p\kern -5.2pt\raise 0.2ex\hbox {/}}

\newcommand{\vdir}{v\kern -5.75pt\raise 0.15ex\hbox {/}}
\newcommand{\kdir}{k\kern -5.75pt\raise 0.15ex\hbox {/}}
\newcommand{\epsdir}{\epsilon\kern -5.0pt\raise 0.15ex\hbox {/}}
\newcommand{\bvdir}{\bar{v}\kern -5.75pt\raise 0.15ex\hbox {/}}
\newcommand{\Ddir}{D\kern -7.75pt\raise 0.20ex\hbox {/}}
\newcommand{\Adir}{A\kern -7.75pt\raise 0.20ex\hbox {/}}
\newcommand{\ldir}{l\kern -5.0pt\raise 0.2ex\hbox{/}}
\newcommand{\varepsdir}{\varepsilon\kern -5.5pt\raise 0.15ex\hbox{/}}

\newcommand{\nn}{\nonumber}

\title{Testing the SM in $B \to D \tau \bar \nu$ decay with minimal theory input}

\ShortTitle{$B \to D \tau \bar\nu$ with minimal
  theory input}

\author{\speaker{Nejc Ko\v snik}\\
        Jo\v zef Stefan Institute, Jamova 39, P. O. Box 3000, 1001 Ljubljana, Slovenia\\
        E-mail: \email{nejc.kosnik@ijs.si}}

\author{Damir Be\v cirevi\'c\\
        Laboratoire de Physique Th\'eorique (B\^at. 210)\thanks{Laboratoire de Physique Th\'eorique est une unit\'e mixte de recherche du CNRS, UMR 8627}, Universit\'e
        Paris Sud, F-91405 Orsay-Cedex, France\\
        E-mail: \email{damir.becirevic@th.u-psud.fr}}

\author{Andrey Tayduganov\\
        Department of Physics, Graduate School of Science, Osaka
        University, Japan\\
        E-mail: \email{taydugan@lal.in2p3.fr}}

\abstract{Recent experimental results for the ratio of the branching fractions of $\bar B \to D^{(*)}\tau \nu_\tau$ and $B\to D^{(*)} \mu \nu_\mu$ decays came as a surprise and lead to a discussion of possibility to constraining new physics through these modes. Here we focus on $Br(B \to D \tau \bar\nu_\tau)/Br(B \to D \mu \bar\nu_\mu)$ and argue that the result is consistent with the Standard Model within $2\,\sigma$ and that the test of compatibility of this ratio with the Standard Model can be done experimentally with a minimal theory input. We also show that these two decay channels can provide us with quite good constraints of the new physics couplings.}

\FullConference{Xth Quark Confinement and the Hadron Spectrum,\\
		October 8-12, 2012\\
		TUM Campus Garching, Munich, Germany}

\begin{document}

\section{Introduction}
Recently BaBar collaboration measured the semileptonic branching
fractions $B \to D \tau \bar \nu$ and $B  \to D^* \tau \bar \nu$ that
are above their Standard Model predictions~\cite{babar:2012xj}. The
experiment reports
\begin{equation}
  \label{eq:R}
  R(D)={{\cal B}(\bar B \to D\tau\bar \nu_\tau) \over {\cal B}(\bar B \to D\mu\bar \nu_\mu) }=0.440\pm 0.058_\mathrm{stat.}\pm 0.042_\mathrm{syst.}\,,
\end{equation}
for the decay with pseudoscalar $D$ and the result was normalized with respect
to the decay with light lepton in the final state in order to cancel
$V_{cb}$ and form factor parameterization in the theoretical
prediction of this observable. Before this result
was published a prediction based on the lattice-calculated form
factors had been made~\cite{prediction}\,,
\begin{equation}
  \label{eq:2}
  R(D)_\mathrm{SM} = 0.296 \pm 0.016\,.
\end{equation}
The discrepancy immediately raised interest in the flavor community
to explain it within one of the well-motivated NP models. On the one
hand, the SM charged-current contribution to this decay hints that
possible NP contributions should be present at tree-level. On the
other hand, the SM prediction requires the $B \to D$ form factors
whose knowledge from the lattice is limited to high-$q^2$ region where
phase space is small.

In this work we have revisited the theoretical prediction of $B \to D
\tau \bar \nu$ in the SM in a manner that maximally employs the
available experimental information on the form factors from previously
measured $B \to D \ell \bar \nu$ where $\ell$ stands for
$e,\mu$~\cite{Becirevic:2012jf}. Next, we parameterize beyond the
Standard Model contributions to this decay mode in the effective
Hamiltonian language, focusing on interactions that preserve lepton
flavor universality and induce $b \to c$ transitions via scalar and
tensor interactions. Assuming presence of either scalar either tensor
operator we derive bounds on their respective Wilson coefficients at
1- and 2-$\sigma$ level, taking into account experimental and
theoretical uncertainties.

\section{Differential decay width}
In the SM the amplitude for hadronic transition $D \to P$ is given in
terms of vector and scalar form factors, $F_+(q^2)$ and $F_0(q^2)$,
defined as
\begin{equation}
  \label{eq:4}
\langle D(p^\prime) \vert \bar c \gamma_\mu b \vert\bar B(p)\rangle  =
\left( p_\mu+p^\prime_\mu - {m_B^2 - m_D^2 \over q^2 } q_\mu \right) F_+(q^2)
+ {m_B^2 - m_D^2 \over q^2 } q_\mu 
 F_0(q^2)\,.  
\end{equation}
The momentum transfer is denoted by $q=p-p^\prime$ and when squared
coincides with the invariant mass of the leptons, $q^2 = (p_{\bar\nu}
+ p_\ell)^2$. In the SM, the differential width of $B \to D \ell \bar \nu$ decay
valid for finite lepton mass $m_\ell$ is
\begin{eqnarray}
\label{eq:dGammadqq}
{d{\cal B}(\bar B \to D\ell \bar \nu_\ell)\over dq^2}&=& \tau_{B^0}{G_F^2\vert V_{cb}\vert^2\over 192 \pi^3 m_B^3} \biggl[ c_+^\ell (q^2) \vert F_+(q^2)\vert^2 +  c_0^\ell  (q^2) \vert F_0(q^2)\vert^2\biggr]\nn\\
&=& \vert V_{cb}\vert^2 {\cal B}_0  \vert F_+(q^2)\vert^2 \left[ c_+^\ell (q^2)  + 
  c_0^\ell (q^2) \left| { F_0(q^2)\over   F_+(q^2)}\right|^2 \right]\,,
\end{eqnarray}
where
\begin{eqnarray}
\label{eq:cs}  
&&c_+^\ell (q^2) =\lambda^{3/2}(q^2,m_B^2,m_D^2) \left[ 1- \frac{3}{2}\frac{ m_\ell^2}{q^2} +  \frac{1}{2}\left(\frac{ m_\ell^2}{q^2}\right)^3\right]\,,\nn\\
&&c_0^\ell (q^2) =m_\ell^2 \ \lambda^{1/2}(q^2,m_B^2,m_D^2) \frac{3 }{2} \frac{ m_B^4}{q^2}\left(1-\frac{ m_\ell^2}{q^2}\right)^2\left(1-\frac{ m_D^2}{m_B^2}\right)^2\,,\nn\\
&&\hfill \nn\\
&& \lambda(q^2,m_B^2,m_D^2) =[q^2 - (m_B+m_D)^2] [q^2 - (m_B-m_D)^2] \,.
\end{eqnarray}
The overall factor ${\cal B}_0$ is defined by
Eq.~(\ref{eq:dGammadqq}):
\begin{equation}
  \label{eq:3}
   {\cal B}_0 = \frac{G_F^2}{192\pi^3 m_B^3} \tau_B\,.
\end{equation}
 The scalar form factor $F_0$ is multiplied by $c_0^\ell(q^2)$ that is
 nonzero only when the lepton in the final state is massive.
Therefore only the $B \to D \tau \bar \nu$ mode is sensitive to
exchanges of charged Higgs in the context of two Higgs doublet models
or alternative scenarios that induce scalar
operators~\cite{fred-jernej,stephanie,alternatives1,alternatives2}.

\begin{figure}
  \begin{tabular}{cc}
    \includegraphics[width=0.48\textwidth]{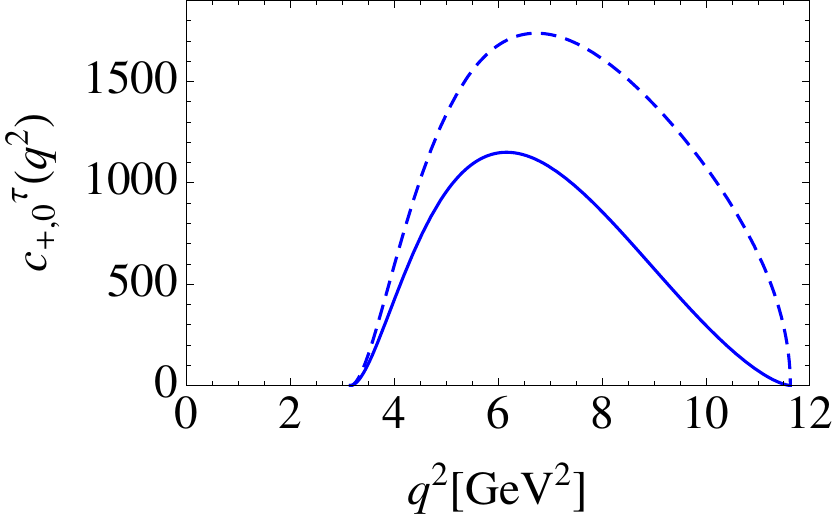} &\includegraphics[width=0.5\textwidth]{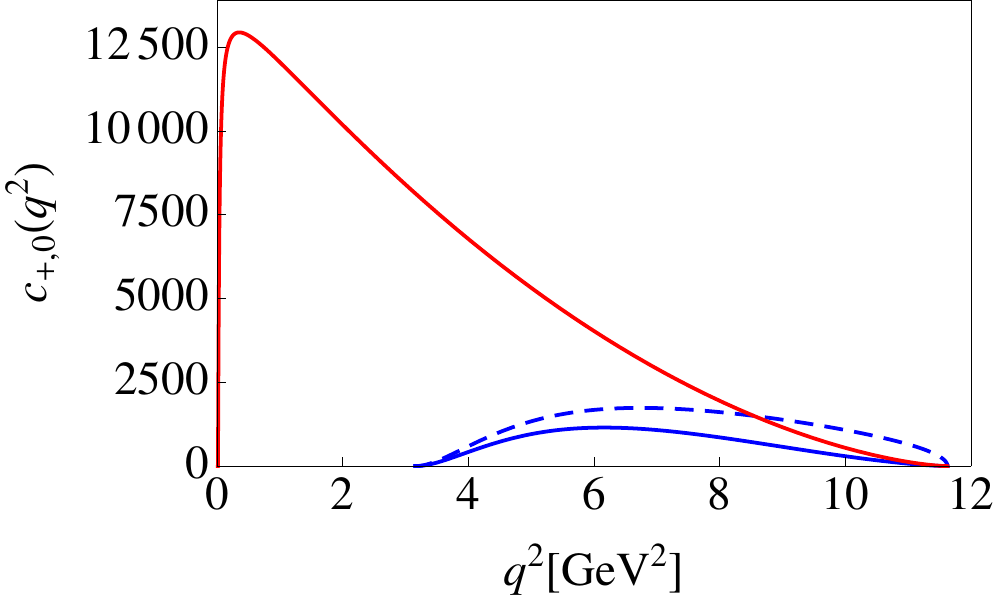}
  \end{tabular}
  \caption{Vector and scalar form factor weight functions for decay $\bar B\to D\tau\bar \nu_\tau$
    are shown in the left plot. Dashed curve corresponds to
    $c_0^\tau(q^2)$, while the solid curve corresponds to
    $c_+^\tau(q^2)$. In the right plot the same
    $c_{+,0}^\tau(q^2)$ are plotted together with $c_+^\mu(q^2)$ (red
    curve), the weight function in the case of massless muon in
    the final state.}
	\label{fig:cdep}
\end{figure}

The $q^2$-dependence of the functions $c^\ell_+(q^2)$ and
$c^\ell_0(q^2)$ are shown in Fig.~\ref{fig:cdep}.
On the left-hand side plot one observes that the contribution of
scalar form factor is enhanced by large $c_0^\tau(q^2)$ that partially
compensates smallness of $F_0(q^2)$.

\section{$B \to D \ell \bar \nu$ with minimal theory input}
In this section we demonstrate, how one can maximally employ available
experimental data to test for consistency of the SM in observable
$R(D)$.  Semileptonic decays with light lepton ($\ell = e, \mu$) have
enabled extraction of $|V_{cb}|$. Usual approach in the literature has
been to make the SM theoretical prediction of the $q^2$-spectrum at
the maximal lepton recoil point $q^2_\mrm{max} = (m_B - m_D)^2$, where
the two hadrons are both at rest, and where the corrections to the
heavy quark limit have been calculated. On the other hand, the
experimental data close to $q^2_\mrm{max}$ have virtually useless
statistics due to small phase space, and experiments rely on the CLN
shape of the form factor $F_+(q^2)$ to fit the differential decay
spectrum to data at low to moderate $q^2$ and extrapolate it to
$q^2_\mrm{max}$~\cite{CLN,Aubert:2009ac}.

\begin{figure}
  \begin{tabular}{cc}
    \includegraphics[width=0.48\textwidth]{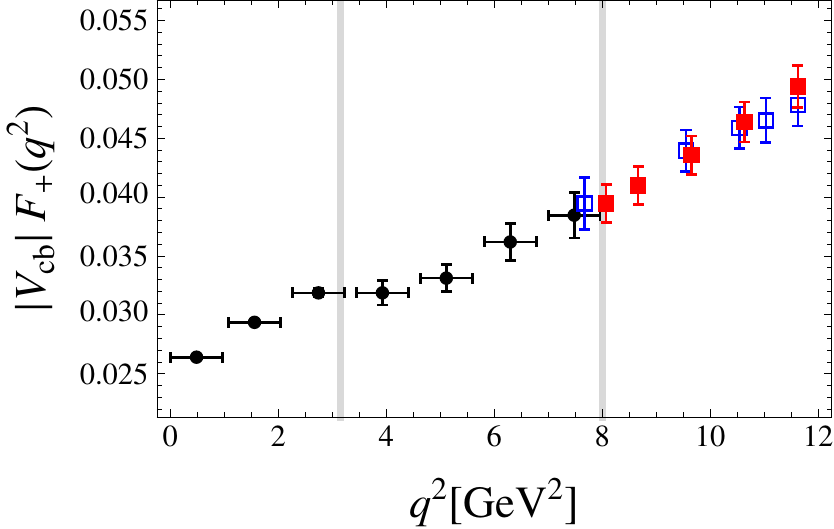} & \includegraphics[width=0.48\textwidth]{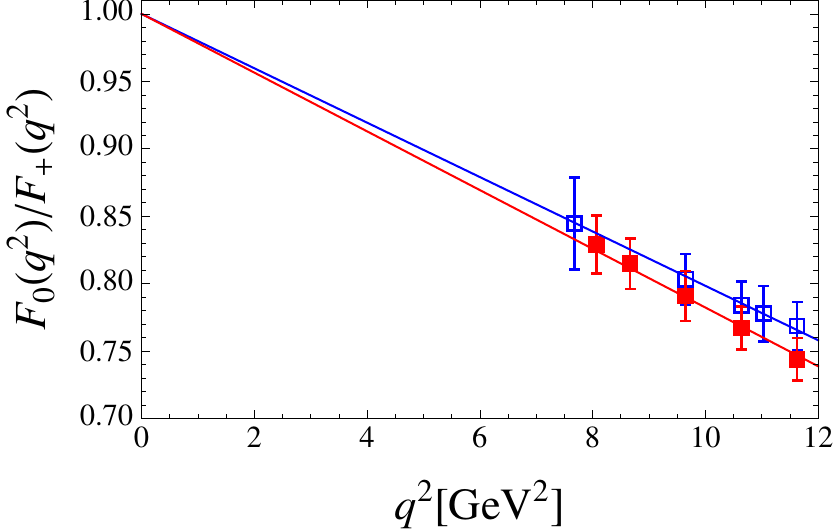}
  \end{tabular}
  \caption{In the left plot we show the binned spectrum of $\bar B\to
    D\mu\bar \nu$ below $8\,\mrm{GeV}^2$ (right vertical grey line) as measured by the BaBar
    collaboration ~\cite{Aubert:2009ac}. Above $8\,\mrm{GeV}^2$
    lattice QCD simulations in quenched approximatian (empty
    symbols)~\cite{nazario}, and those in which $N_{\rm f}=2+1$
    dynamical flavors are included (filled
    symbols)~\cite{Bailey:2012rr} are indicated. First grey line on
    the left denote the $\tau$ threshold at $q^2 = m_\tau^2$. Ratio of $B\to D$
    form factors obtained in the same lattice simulations are shown on
    the right plot.}
	\label{fig:ff}
\end{figure}

In order to maximally use the experimental data we propose a different
path for both light and heavy leptons in the final state of $B \to D
\ell \bar \nu$. The vector form factor can be extracted from the BaBar
measurement of the spectrum with light leptons in the final
state. This procedure is viable below $q^2 = 8\,\mrm{GeV}^2$, where
the errors in bins are small~\cite{Aubert:2009ac}. Prediction of $B
\to D \tau \bar \nu$ in $q^2 \in [8\,\mrm{GeV}^2, q^2_\mrm{max}]$ region
will rely on the lattice QCD results for the vector form
factor~\cite{nazario,Bailey:2012rr} as seen in the left panel of
Fig.~\ref{fig:ff}.
Notice that below the above procedure relies on the
input value of $V_{cb}$ only at high $q^2$ where lattice data is
used. 

Now we can express both branching fraction entering $R(D)$ as integrals over
the three kinematical regions denoted in Fig.~\ref{fig:ff}
\begin{eqnarray}
{\cal B}(\bar B \to D\mu\bar \nu_\mu)
&=&
 {\cal B}_0 \int_{m_\mu^2}^{8\,\mathrm{GeV}^2}c_+^\mu (q^2)   \vert V_{cb}
 F_+(q^2)\vert_\mrm{exp}^2 dq^2 \nn \\
&&+\vert V_{cb}\vert^2 {\cal B}_0 \int_{8\,\mathrm{GeV}^2}^{q_{\rm max}^2}c_+^\mu (q^2)   \vert F_{+,\mrm{latt}}(q^2)\vert^2 dq^2 \,,
\end{eqnarray}
where $q^2_{\rm max}=(m_B-m_D)^2$. The phase space
integral for decay with heavy $\tau$ contains the contribution of the
scalar form factor and is also split at $q^2 =
8\,\mathrm{GeV}^2$
\begin{eqnarray}
{\cal B}(\bar B \to D\tau\bar \nu_\tau)
&=&
 {\cal B}_0 \int_{m_\tau^2}^{8\,\mathrm{GeV}^2}   \vert
 V_{cb}  F_+(q^2)\vert_\mathrm{exp}^2 \left[ c_+^\tau (q^2) +  c_0^\tau (q^2)
   \left| {F_0(q^2)\over F_+(q^2)}\right|^2 \right]dq^2 \nn\\
 &&+
   \vert V_{cb}\vert^2 {\cal B}_0 \int_{8\,\mathrm{GeV}^2}^{q_{\rm max}^2}   \vert F_{+,\mathrm{latt}}(q^2)\vert^2  \left[ c_+^\tau (q^2) +  c_0^\tau  (q^2) \left| { F_0(q^2)\over   F_+(q^2)}\right|^2 \right] dq^2\,.
\end{eqnarray}
We emphasize again that the partial decay rates at low-$q^2$ do not
rely on the input value of $V_{cb}$, since $|V_{cb} F_+(q^2)|$ is
available experimentally. A suitable observable with this pleasing
property should be defined as a ratio of partial decay rates, i.e.,
\begin{equation}
  \label{eq:5}
  R(D)\big|_{q^2 \leq 8\,\mathrm{GeV}^2} = \left. {{\cal B}(\bar B \to D\tau\bar \nu_\tau) \over {\cal B}(\bar B \to D\mu\bar \nu_\mu) }\right|_{q^2\leq 8\ \gev^2}\,,
\end{equation}
and would depend only on $|F_0/F_+|$ that is well
under control theoretically as we show below. The theoretical
prediction of $R(D)$, as defined in Eq.~(\ref{eq:R}), requires in
addition the SM value of $V_{cb}$ and calculation of the vector form
factor at high $q^2$. For the latter we take the lattice
results of Refs.~\cite{nazario,Bailey:2012rr} and fit them to a
dipole parameterization. The global fits of the SM to flavor
observables yield $|V_{cb}| = 0.0411(16)$, a value that we use
above $8\,\mathrm{GeV}^2$~\cite{ckm}.

The ratio of scalar-to-vector form factor is constrained to be $1$ at
$q^2 = 0$ by construction and indicates, together with high $q^2$
results of lattice QCD, a linear behaviour (see Fig.~\ref{fig:ff})
\begin{equation}
  \label{eq:f0f+}
  {F_0(q^2)\over F_+(q^2)} = 1 - \alpha\ q^2\,.
\end{equation}
We use the value $\alpha = 0.020(1)\,\mathrm{GeV}^{-2}$ that is
consistent with different theoretical approaches (see
e.g.~\cite{nazario,Bailey:2012rr,Melikhov:2000yu,Faller:2008tr,Azizi:2008tt}). The
above procedure gives finally
\begin{equation}
  \label{eq:our-value}
  R(D)=0.31\pm 0.02\,,
\end{equation}
which is less than $2\sigma$ below the BaBar result~(\ref{eq:R}).

\section{Constraint on NP effective Hamiltonian}
The SM effective Hamiltonian is extended to include operators that are
scalars, tensors, or vectors, their dimensionless couplings labelled as $g_S$,
$g_T$, and $g_V$, respectively. We assume that the lepton flavor
universality is respected by all couplings on the Lagrangian level.
We do not invoke operators that contain a right-handed neutrino. 
\begin{eqnarray}\label{eq:lagr2}
{\cal H}_{\rm eff}=  -{ \sqrt{2} G_F V_{cb}}&&\left[( \bar c \gamma_\mu b)( \bar \ell_L \gamma^\mu\nu_L)  + g_V (\bar c \gamma_\mu b)( \bar \ell_L \gamma^\mu\nu_L)  \right.\nn\\
&&\left. + g_S(\mu)( \bar c b )(\bar \ell_R \nu_L) 
+ g_T(\mu)( \bar c \sigma_{\mu \nu}b )(\bar \ell_R \sigma^{\mu \nu} \nu_L)\right] + {\rm h.c.} 
\end{eqnarray}
All couplings scale as $ g_{V,S,T}\propto m_W^2/m_{\rm NP}^2$, and
$m_{\rm NP}$ is the new physics scale. The differential decay width
in the presence of NP operators is 
\begin{eqnarray}\label{eq:0}
{d{\cal B}(\bar B \to D\ell \bar \nu_\ell)\over dq^2} = \vert V_{cb}\vert^2 &{\cal B}_0&  \vert F_+(q^2)\vert^2 \left\{ \vert 1 + g_V\vert^2 c_+^\ell (q^2) + \vert g_T(\mu)\vert^2 c_T^\ell (q^2 )   \left|{F_T(q^2,\mu)\over F_+(q^2)}\right|^2\right. \nn\\
&+&    c_{TV}^\ell (q^2 ) \ {\rm Re} \left[(1+g_V) g_T^\ast (\mu)  \ {F_T(q^2,\mu)\over F_+(q^2)} \right] \nn\\
&+&\left.    \left| (1+g_V) -  {q^2\over m_\ell}{g_S(\mu) \over m_b(\mu)-m_c(\mu)} \right|^2c_0^\ell (q^2) \left| { F_0(q^2)\over   F_+(q^2)}\right|^2 \right\}\,,
\end{eqnarray}
where
\begin{eqnarray}
c_T^\ell (q^2,\mu)&=& \lambda^{3/2}(q^2,m_B^2,m_D^2) {2  q^2 \over (m_B+m_D)^2}  \left[1- 3 \left(\frac{ m_\ell^2}{q^2}\right)^2 +2 \left(\frac{ m_\ell^2}{q^2}  \right)^3\right] \,,\nn\\
c_{TV}^\ell (q^2,\mu) &=&  {6 m_\ell \over   m_B+m_D } \  \lambda^{3/2}(q^2,m_B^2,m_D^2) \left(1-  \frac{ m_\ell^2}{q^2} \right)^2\,.
\end{eqnarray}
The tensor form factor $F_T(q^2,\mu)$ is defined as
\bea\label{eq:fTT}
\langle D(p^\prime) \vert \bar c \sigma_{\mu\nu} b \vert\bar B(p)\rangle  = - i
\left( p_\mu p^\prime_\nu -  p^\prime_\mu p_\nu \right) {2 \ F_T(q^2,\mu) \over m_B+m_D} \;.
\eea

While the vector nature of weak currents has been thouroughly tested
and are compatible with $g_V = 0$, scalar $g_S(\mu)$ and tensor
$g_T(\mu)$ operators are not as constrained. $g_S(\mu)\neq 0$ induces
the left-right operator and lifts the helicity suppression in $\bar
B\to D\tau \bar \nu$ and $\bar B\to D\mu \bar \nu$ decays. Same
operator is enhanced by the factor $m_D^2/m_c^2$ with respect to the
left-left (SM) contribution to the $D^0-\bar D^0$ mixing amplitude.
Furthermore a noticable effect could also be seen in $D\to V\gamma$
decays that are governed by loops containing the down-type
quarks and are therefore sensitive to
$g_S(\mu)\neq 0$~\cite{jernej-gino}.

\begin{figure}
  \begin{tabular}{cc}
    \includegraphics[width=\textwidth*\real{0.98}*\real{0.5}]{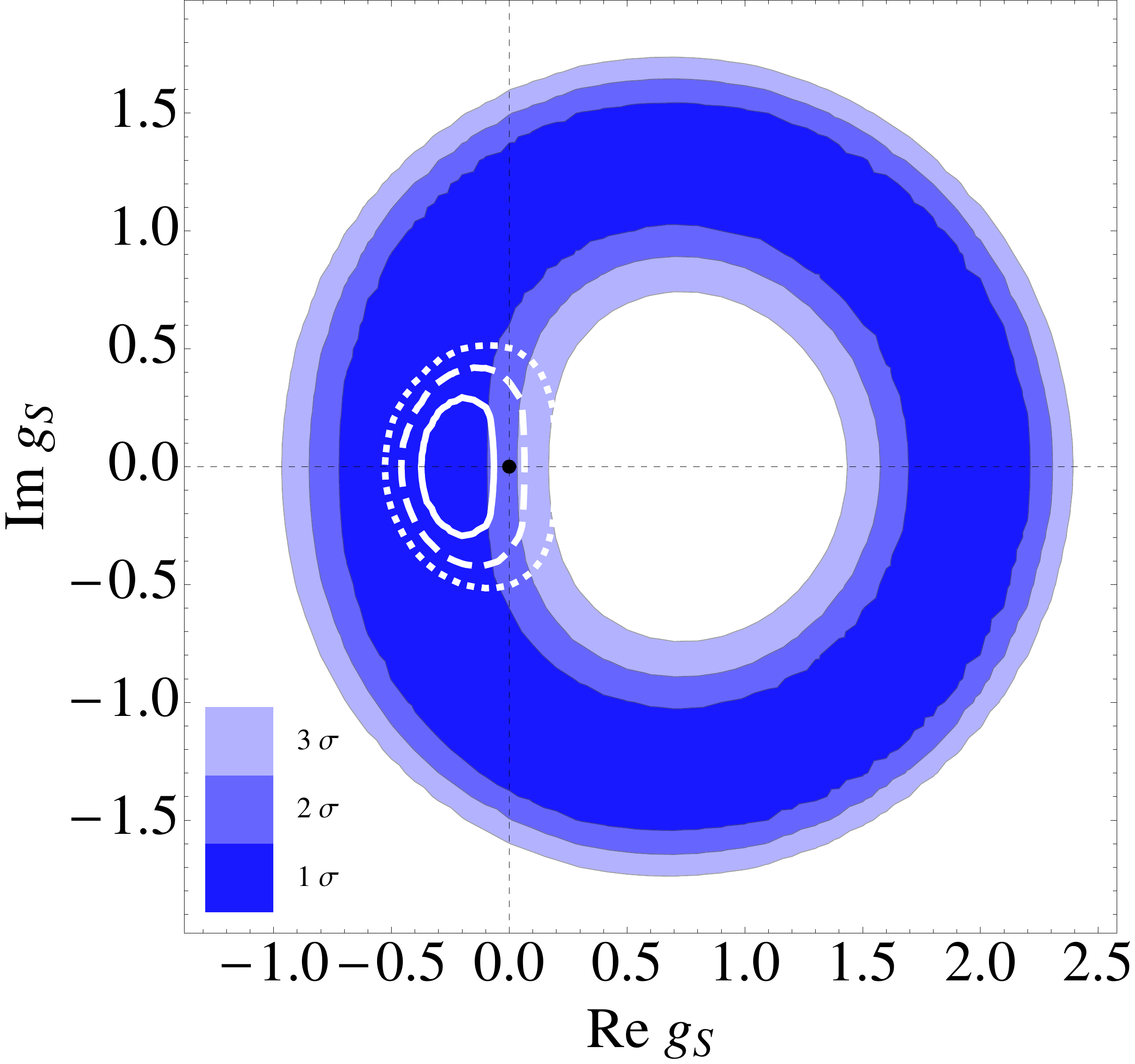} & \includegraphics[width=\textwidth*\real{0.98}*\real{0.474}]{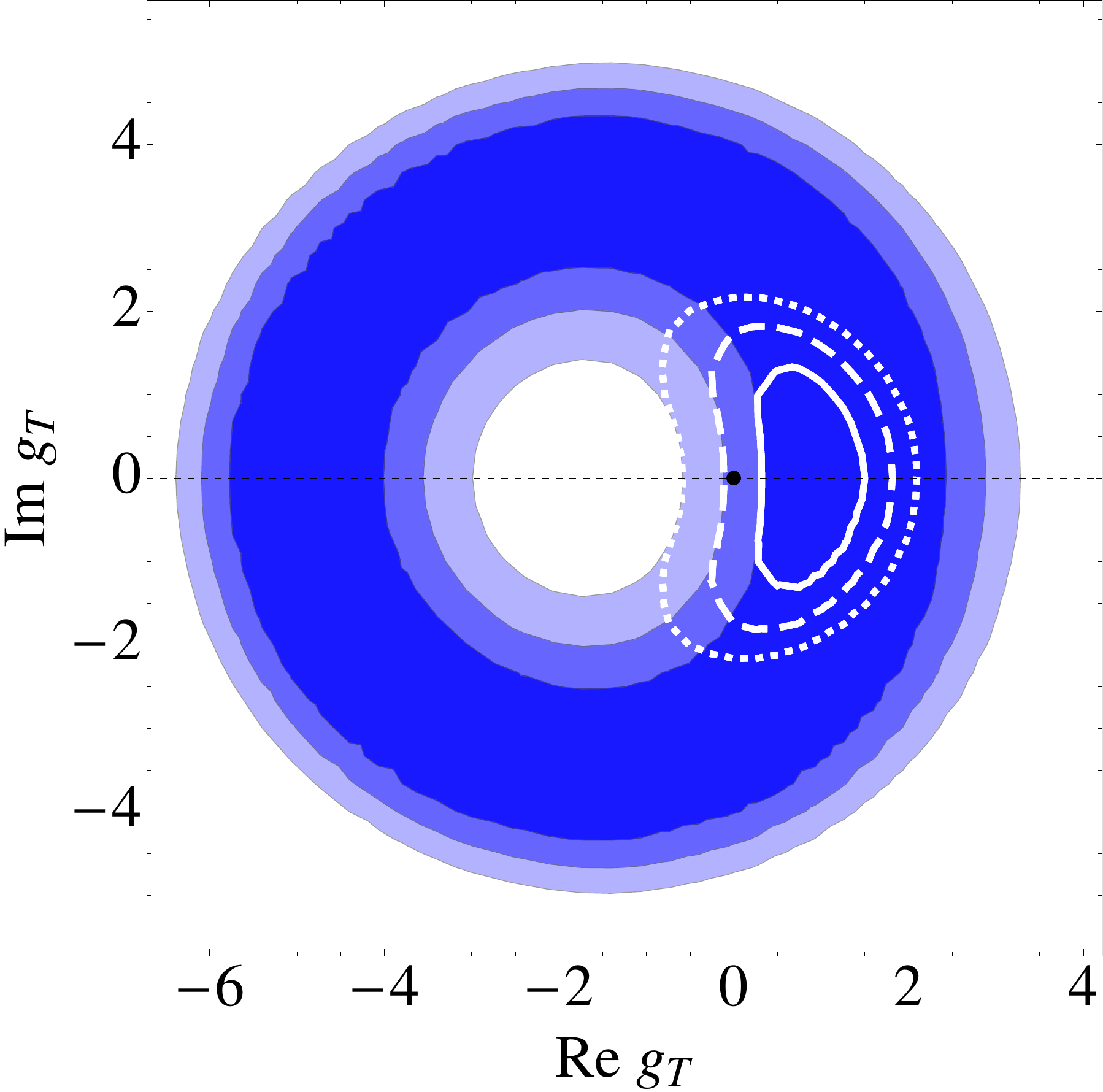}
  \end{tabular}
  \caption{Regions of allowed values for $g_S(m_b)$ and $g_T(m_b)$,
    compatible with experimentally measured $R(D)$. The small region
    within the solid, dashed and dot-dashed white curves correspond to
    the respective $1$-, $2$- and $3$-$\sigma$ compatibility with both
    ${\cal B}^{\rm (exp)}(\bar B\to D\tau\bar \nu_\tau)$ and ${\cal
      B}^{\rm (exp)}(\bar B\to D\mu\bar \nu_\mu)$. The thick dot
    represents the Standard Model, namely $g_{S,T}(m_b)=0$.}
\label{fig:gSgT}
\end{figure}
Using $R(D)$ alone we get a very loose constraint on $g_S(m_b)$ while
we require $g_V=g_T(m_b)=0$ (contours on the left-hand plot in
Fig.~\ref{fig:gSgT}). Requiring in addition the compatibility of the
theoretical expression for ${\cal B}(\bar B\to D\mu\bar \nu_\mu)$
obtained by using Eq.~(\ref{eq:0}) and the measured
value~\cite{Aubert:2009ac}, restricts the allowed $g_S(m_b)$ to a
small region also indicated in Fig.~\ref{fig:gSgT}.  For example, when
$g_S(m_b)$ is real then the $1\sigma$ compatibility with experiment
allows $-0.37\leq g_S(m_b) \leq -0.05$, while the $3\sigma$
compatibility amounts to $-0.53\leq g_S(m_b) \leq +0.20$. The
statistical error in Eq.~(\ref{eq:R}) is treated as Gaussian, while
the systematic errors and uncertainties with respect to the form
factors are treated as uniform.

If we allow for $g_T(m_b)\neq 0$ in Eq.~(\ref{eq:0}) then the possible
values that are compatible with $R(D)$ are those
in the contour plot shown in the right-hand plot of
Fig.~\ref{fig:gSgT}. The needed tensor form factor has not been
computed, to our knowledge, on the lattice nor in the QCD sum
rules. The computation in the model of ref.~\cite{Melikhov:2000yu}
shows that $F_T(q^2)/F_+(q^2) = 1.03(1)$ is a constant, in agreement
with naive expectations based on the pole dominance. As before,
$R(D)$ alone is not constraining strongly the possible
values of $g_T(m_b)$ whereas taking into account the constraint from measured ${\cal
  B}(\bar B\to D\mu\bar \nu_\mu)$~\cite{Aubert:2009ac} shrinks the
allowed region, as shown in the right-hand plot in
Fig.~\ref{fig:gSgT}. If ${\rm Im}\ g_T(m_b)=0$, we obtain $0.3\leq
g_T(m_b) \leq 1.5$ and $-0.6\leq g_T(m_b) \leq 2.1$, from the
requirement of respective $1$- and $3\sigma$ compatibility with both
experimental $R(D)$ and ${\cal B}(\bar B\to D\mu\bar \nu_\mu)$. 

\section{Conclusions}

  The result for $R(D)$ could be an indication of new physics should
  the significance of incompatibility with the Standard Model raise to
  at least $3\sigma$. The compatibility with the Standard Model can be
  tested experimentally, with a minimal hadronic input, as discussed
  in this letter.  Here we used the lattice QCD results for $F_+(q^2)$
  at larger $q^2$'s because the full branching fractions were reported
  in Ref.~\cite{babar:2012xj}. Thus we have found $R(D) = 0.31 \pm
  0.02$ where the significance of the discrepancy with
  Eq.~(\ref{eq:R}) to be below 2$\sigma$.

  If, instead of comparing the full branching fractions of both decay
  modes, the experimenters cut at about $q^2\approx 8~\gev^2$, then
  the shape of the needed vector form factor including the factor of
  $|V_{cb}|$ could be reconstructed from the differential branching
  fraction of $\bar B\to D\mu\bar
  \nu$~\cite{fred-jernej,stephanie}. The only theoretical hadronic
  quantity needed then is the slope of the form factor
  ratio~(\ref{eq:f0f+}), which is quite accurately known from lattice
  QCD with the values that agree with quark models and with recent QCD
  sum rule studies. By using the vector form factor multiplied by $|V_{cb}|$ data from
  Ref.~\cite{Aubert:2009ac} only, and by integrating the decay rates
  up to $q^2_{\rm cut}=8\ \gev^2$, we obtain \bea \left. {{\cal
        B}(\bar B \to D\tau\bar \nu_\tau) \over {\cal B}(\bar B \to
      D\mu\bar \nu_\mu) }\right|_{q^2\leq 8\ \gev^2} =0.20\pm 0.02\,.
  \eea

  Allowing for departures from the Standard Model, while keeping
  lepton flavor universality which has been experimentally verified
  to a very good accuracy~\cite{PDG}, the measured $R(D)$ and ${\cal
    B}(\bar B\to D\mu\bar \nu_\mu)$ impose quite strong constraints on
  the new physics scalar and tensor effective 
  couplings $g_{S,T}(m_b)$.

\end{document}